\documentclass[dvips]{article}
\usepackage{subfigure}
\usepackage{amsmath}
\usepackage{icrctc07}

\newcommand{\be}{\begin{equation}}
\newcommand{\ee}{\end{equation}}
\newcommand{\ben}{\begin{eqnarray}}
\newcommand{\een}{\end{eqnarray}}
\newcommand{\bc}{\begin{center}}
\newcommand{\ec}{\end{center}}

\title{The Active Mirror Control of the MAGIC Telescope}
\authors{A. Biland$^{1}$, 
M. Garczarczyk$^{2}$,
H. Anderhub$^{1}$,
V. Danielyan$^{3}$,
D. Hakobyan$^{3}$,
E. Lorenz$^{1}$,
R. Mirzoyan$^{2}$
on behalf of the MAGIC
collaboration$^{*}$}

\shortauthors{A. Biland et al.}

\afiliations{$^1$ETH Zurich, Institute for Particle Physics, 8093 Zurich, Switzerland\\
  $^2$Max-Planck-Institut f\"ur Physik, F\"ohringer Ring 6, 80805 M\"uchen, Germany\\
  $^3$Yerevan Physics Institute, Cosmic Ray Division, Yerevan 375022, Armenia\\
  $^*$See \texttt{http://wwwmagic.mppmu.mpg.de/collaboration/members/}}

\email{biland@phys.ethz.ch, garcz@mppmu.mpg.de}

\abstract{One of the main design goals of the MAGIC telescopes is the very fast 
repositioning in case of Gamma Ray Burst (GRB)
alarms, implying a low weight of the telescope
dish. This is accomplished by using a space frame made of carbon fiber epoxy tubes,
resulting in a strong but not very rigid support structure. Therefore it is
necessary to readjust the individual mirror tiles to correct for deformations of
the dish under varying gravitational load while tracking an object.
We present the concept of the Active Mirror Control (AMC) as implemented in the
MAGIC telescopes and the actual performance reached.
Additionally we show that also telescopes using a stiff structure can benefit from
using an AMC.}

\begin{document}

\maketitle
\section{Introduction}
\vspace*{-1mm}
The MAGIC telescope~\cite{MGC} is installed at the Roque de los
Muchachos on a height of $\approx$2200 m a.s.l. on the Canary Island 
La Palma. Having a reflector area of 236 m$^2$, it is by far the 
largest existing imaging Cherenkov telescope.
One of the main goals in the design of the MAGIC telescope 
was the ability to point to any arbitrary position on the sky in 
less than 20 s to allow e.g.~the observation of GRB
immediately after an alert from satellite
experiments. To achieve this goal, a stiff, lightweight frame 
structure made of carbon fiber epoxy tubes was built to minimize the
telescope's weight. When pointing the telescope to different 
zenith angles, the reflector's surface undergoes small deformations
under varying gravitation loads. To correct for 
this effect, the individual mirror-segments must be 
realigned~\cite{AMC1,AMC2}.

\section{Implementation}
\vspace*{-1mm}
The reflector of the MAGIC telescope consists of 956 square 
shaped mirrors with a size of 49.5 x 49.5 cm$^2$. Four (at the 
edges three) mirrors are mounted and adjusted on one of 247 
panels.

\begin{figure}[!hb] \centering
   \includegraphics[scale=.42]{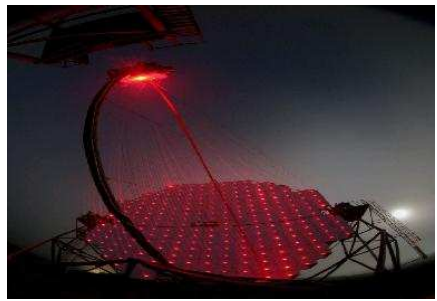}
   \caption{MAGIC telescope with all AMC guidance lasers switched on.}
   \label{fig:Biland-fig0}
\end{figure}

These panels are attached to the carbon fiber support 
frame at three points; one of these points has a fixed distance,
while the other two are equipped with longitudinally movable
actuators having a lateral 
freedom of one and two dimensions respectively 
Each actuator
contains a two phase stepping motor and a ball-bearing-spindle,
allowing to move the panels with a precision $<20 \mu$m,
corresponding to a displacement of the light spot at the 
camera of less than 1 mm. To monitor the orientation of a
panel, each one is equipped with a guidance
laser module, pre-adjusted 
to point to the target of the panels
 (figure \ref{fig:Biland-fig0}).

\begin{figure}[!ht] \centering
   \includegraphics[scale=.24]{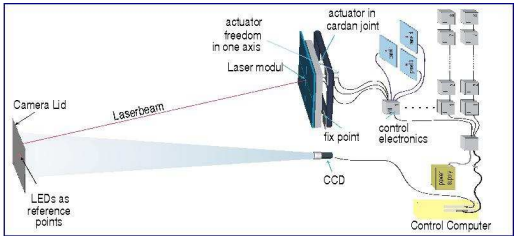}
   \caption{Components of the Active Mirror Control.
    See text for details.}
   \label{fig:Biland-fig1}
\end{figure}

As illustrated in figure \ref{fig:Biland-fig1}, the
AMC also includes four LEDs on the PMT-camera lid as 
reference points and a CCD-camera mounted on the reflector 
frame close to its center. The electronics to handle the 
actuators and lasers is housed in 62 water tight boxes and 
the connection to the control computer is split into eight 
independent RS-485 strings to allow for parallel operation.

\section{Initial Mirror Alignment}
In a first step it is needed to calibrate the full system, 
i.e.~to find the correlation between the movements of the stepping motors
and the displacement of the reflected light. This is done by measuring
the position of the laser spot for several stepping motor positions
for each panel.

Next, one has to find the best mirror alignment for one telescope
orientation. Originally, this was done
using a bright, artificial point-like light source
pointing towards the telescope and adjusting each
individual panel manually
so that the light is reflected to the center of 
the PMT camera. Then the camera-lid was closed, each laser
switched on sequentially and the position of the laser spot
on the lid
relative to the LEDs as measured by the CCD is stored as
reference point. 
Unfortunately, the farthest possible place to
mount the artificial light source is only 980 m away from MAGIC.
While for a spherical reflector it would be sufficient to move the PMT-camera 
27 cm closer to the dish to change the focal length from 980 m to 10 km, 
this is not enough in the case of a paraboloid: the outermost panels 
would still point as much as 25 mm off target. Therefore, a higher
order correction had to be
be calculated and applied to these reference points. An additional problem
was that the (temporary) mounting of the light-source was not exactly
reproducible and therefore existed the risk to introduce an artifical
miss-pointing of the telescope.

\begin{figure}[!h] \centering
   \includegraphics[scale=.48]{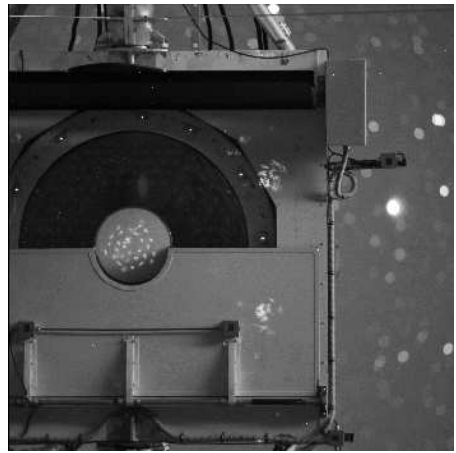}
   \caption{Light-spots of 30 mirror tiles defocused to a pre-calculated
    pattern on a reflector at the center of the PMT camera.}
   \label{fig:Biland-fig2}
\end{figure}

To overcome this restriction, the CCD-camera was upgraded to higher
sensitivity to be able to see the reflection of a single panel while
tracking a bright star. Additionally, a mechanism to position a
highly reflective panel in front of the PMT camera was installed.
By completely defocusing all but one panel,
it is therefore possible to find the best alignment. But since this
procedure needs several hours and the stars visible under small
zenith angles move rather fast and have to be tracked by the
telescope, this does not result in a focusing for
one fixed zenith angle. Therefore, a method was developed to defocus
about 30 panels according to a pre-calculated grid arrangement (and
completely defocus the remaining $\approx$210 panels) as shown in
figure \ref{fig:Biland-fig2}. As long as the displacement of the light
spot of a panel is less than $\approx$3 cm, it can be identified and
a correction be calculated and applied. Since this needs only 8
measurements to control the alignment of the 247 panels it can be completed
in few minutes and the movement of the telescope can be neglected.

After knowing the best alignment of each panel for one given zenith angle,
the position of the spot of the guidance laser for each panel is stored in
a database and used to transpond the alignment to different zenith angles.

\section{AMC Operation Modes}
In the basic {\em laser adjustment} mode of the AMC, the following steps
have to be done:\\
a) close the camera-lid (to be used as screen for the laser
spots as well as to protect the PMTs from the lasers)\\ 
b) activate the CCD and search for the LEDs that define the
center of the camera \\
c) sequentially for each panel: activate the laser and move
the actuators until the laser position agrees with its
panel's reference point\\ 
d) open the camera-lid and (re)start data-taking\\
This allowes to refocus the whole telescope in about 5 min.
 
A time of 5 min is too slow in case of e.g.~a GRB-alarm or while
tracking a source. 
Therefore it was originally forseen to apply such laser adjustments
only when the tracking of a new source starts, and just apply a
{\em look-up table adjustment} during the tracking. For this,
a laser adjustment is done for many different telescope
orientations and the positions of all stepping motors are stored
in a database. From this database it is possible to extract
the best position of the stepping motors for any telescope
orientation. By calculating the relative change for the stepping
motors between the actual orientation of the telescope and the
orientation when the last laser adjustment was applied, it is
also possible to take into account some time-dependent deformations
of the telescope dish, e.g.~because of temperature changes.
Since this adjustment can be done for several panels in
parallel, it takes less than 10 seconds and can be applied without
interrupting the data-taking. Additionally in case of a fast
repositioning of the telescope like for a GRB-alarm, it can
be done while the telescope is slewing to the new orientation and
the full reflector is well aligned when the telescope is ready
to start data-taking.

\section{Performance}
\begin{figure}[!h] \centering
   \includegraphics[scale=.50]{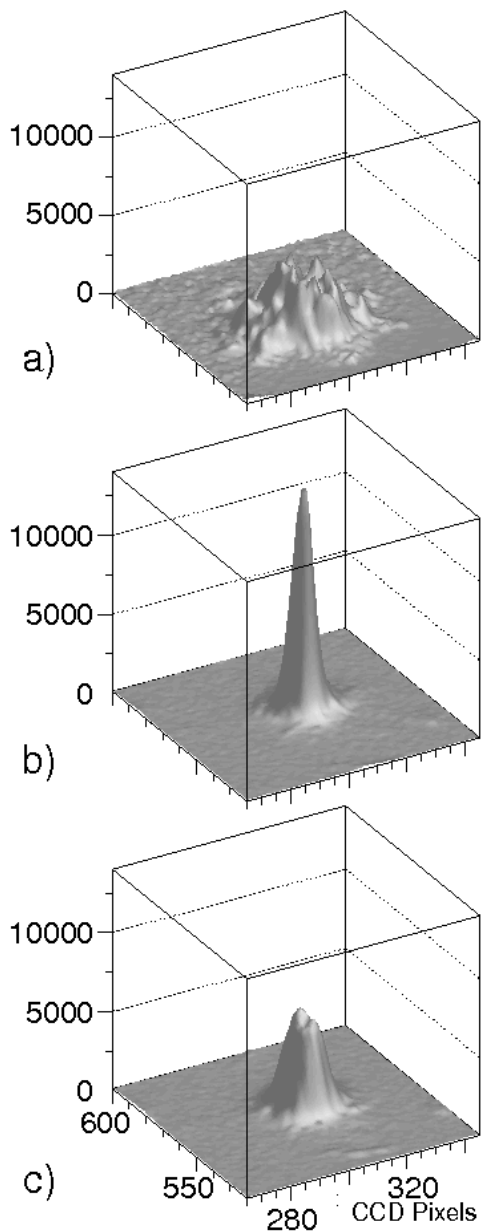}
   \caption{CCD-measurements of a bright star\\
   a) mirror tiles aligned to wrong zenith angle,\\
   b) correct alignment, focal length infinity, \\
   c) correct alignment, focal length 10 km
   }
   \label{fig:Biland-fig3}
\end{figure}

While it is difficult to show the effect of the AMC on individual 
showers, it can easily be seen when pointing to stars. 
Figure \ref{fig:Biland-fig3}a,b
shows the reflected starlight on a
screen in the focus point of the telescope, photographed with 
a high resolution CCD. The top picture is taken with the
panels pre-aligned for a zenith-angle of 80$^\circ$, while the
middle one shows the star after doing a laser-adjustment
in the correct orientation of 22$^\circ$. After adjustment, more
than 85\% of the light is concentrated in an area of 3.5 cm diameter,
corresponding to the size of one central PMT.
The total point spread function is limited by the spot size of
the individual mirror
tiles, while the deformation of the dish is
compensated by the AMC.

Control measurements done with stars on many different telescope
orientations do not only proof the optical quality of the
telescope to be constant over the full zenith range, but also
show no indication for any time- or temperature-dependence of the
deformation of the dish. 
Lookup tables produced in a cold winter night are still giving good
optical alignment even in hot summer. The AMC system of the MAGIC
telescope has been operated in the last year with constant optical
performance without the need of a new calibration.

\section{Additional Benefits}
While an AMC is more expensive than non-active mirror mounts, this can be compensated
by relaxed requirements on the stiffness of the support structure of the dish.

But having the possibility to control each individual panel does not only allow to compensate
deformations of the support structure, but has several other benefits:\\
a) Monitoring the point spread function of a Cherenkov telescope by taking pictures
 from stars is not straight forward: while the star is located in infinity, Cherenkov
 telescopes usually are focusing to the average shower maximum of typically 10--15~km.
 The difference of the images of a bright star using a focal length of 10 km vs.
 infinity can be seen in figure \ref{fig:Biland-fig3}b,c.


\noindent
For spherical reflectors, this can be corrected for by moving the (heavy) PMT camera
by a few centimeters or putting a temporary reflector in front of it.
But for parabolic reflectors, this is not the case because of higher
order corrections that have to be taken into account.
Telescopes of the size of MAGIC (17 m) or larger must have
parabolic reflectors to keep the time dispersion of the Cherenkov signal
short. So
even in case of stiff reflectors, an AMC simplifies the monitoring of
the point spread function. \\
\noindent
b) An AMC can change the focal length to basically any value. Taking into account
that the distance to the shower maximum depends on the zenith angle, an AMC allows
to adjust the focal length during tracking\footnote{This is not done so far in MAGIC, but
will be tested during the commissioning of the second telescope}.\\
\noindent
c) The possibility to easily measure the reflections of a single panel
(or of a group of panels as in figure \ref{fig:Biland-fig2}) allows to monitor
the reflectivity of each individual mirror tile. This allows to identify and
replace aging parts of the reflector.\\
\noindent
d) For daytime, it is possible to defocus all panels so far that the danger of
destroying vital parts of the telescope by accidental exposition to sunlight
is much reduced. This adds another security factor in case of robotic operations of
future telescopes.\\
\noindent
e) It is not only possible to change the focal length of the telescope, but also the
position of the reflected spot can be easily changed. This might allow to use
Cherenkov telescopes as solar power plants during day time without endangering the
valuable PMT camera.

\section{Conclusions}
After some initial problems, the AMC of the MAGIC telescope is working well within the
design limits. This proofs that it is possible to overcome mechanical deformations
of the support structure of the reflector during data taking without any degradation
of the performance of the telescope.
This allows to build light weight structures allowing for a very fast
repositioning and/or relaxing the requirements on the stiffness of the support structure
and reduce therefore the costs.

\end{document}